\documentclass[sigconf, screen, pageno, nonacm]{acmart}
\usepackage{tikz}
\usepackage{graphicx}
\usepackage{xcolor}
\usepackage[]{hyperref}
\usepackage{subcaption}
\usepackage{multirow}
\usepackage{tikz}
\usepackage{amsmath}
\usepackage{xspace}
\usepackage{natbib}
\usepackage{enumitem}
\usepackage{pifont}
\usepackage{xfrac}
\usepackage{balance}
\usepackage{amsthm}
\usepackage{caption}
\usepackage{graphicx}
\usepackage{booktabs}
\usepackage{algorithm}
\usepackage{algpseudocode}
\numberwithin{algorithm}{section}

\begin{document}
\sloppy{
\emergencystretch 3em
}

\newcommand{\iid}{\texttt{iid}\xspace}
\newcommand{\noniid}{non-\texttt{iid}\xspace}
\newcommand{\cselect}{\texttt{CSelect}\xspace}
\newcommand{\cprov}{\texttt{CProv}\xspace}
\newcommand{\cscale}{\texttt{CScale}\xspace}
\newcommand{\emissionunit}{\ensuremath{g\cdot CO_{2}eq}\xspace}
\newcommand{\ciunit}{\ensuremath{g\cdot CO_{2}eq/kWh}\xspace}
\newcommand{\cciunit}{\ensuremath{g\cdot CO_{2}eq/cycle}\xspace}
\newcommand{\efunit}{\ensuremath{cycle/kWh}\xspace}
\newcommand{\xmark}{\ding{55}}
\newcommand{\tickmark}{\ding{51}}
\newcommand{\sysname}{\textsc{PowerScale}\xspace}
\newcommand{\powertrip}{\textsc{PowerTrip}\xspace}

 \newcommand{\talha}[1]{  
	{\textcolor{purple}{(\textbf{Revision:}  #1)}}{}}

\date{}

\title[PowerScale]{PowerScale: Energy-Efficient Geo-Distributed Model Training with Federated Datacenter Power}

\author{Talha Mehboob}
\affiliation{
  \institution{University of Massachusetts Amherst}
  \city{Amherst}
  \state{Massachusetts}
  \country{USA}
}
\email{tmehboob@umass.edu}

\author{Zhe Xu}
\affiliation{
  \institution{University of Massachusetts Amherst}
  \city{Amherst}
  \state{Massachusetts}
  \country{USA}
}
\email{zhexu@umass.edu}

\author{Michael Zink}
\affiliation{
  \institution{University of Massachusetts Amherst}
  \city{Amherst}
  \state{Massachusetts}
  \country{USA}
}
\email{mzink@cas.umass.edu}

\author{David Irwin}
\affiliation{
  \institution{University of Massachusetts Amherst}
  \city{Amherst}
  \state{Massachusetts}
  \country{USA}
}
\email{irwin@ecs.umass.edu}

\begin{abstract}
The power demands of large-scale AI training increasingly exceed the capacity of any single data center, making geo-distributed training across power-constrained sites a practical necessity. Prior work optimizes such training mainly for time-to-accuracy, relying on single-tier aggregation, where every site exchanges model updates directly with a central aggregator over the wide-area network (WAN) each synchronization round, without accounting for the energy required to reach convergence. Single-tier aggregation, however, is fundamentally energy-inefficient for three reasons. First, each round is bottlenecked by the most distant participant, forcing faster sites to draw near-idle GPU power while they wait. Second, every site transmits its full model update over long-haul WAN links, such that beyond a certain site count, communication energy dominates the compute budget. Third, a fixed synchronization frequency pays the same communication cost throughout training, even in later stages where parameter updates shrink and further synchronization yields diminishing benefits.

To address these inefficiencies, we present \sysname, a hierarchical aggregation system that exploits the natural latency hierarchy of wide-area networks. \sysname organizes sites into regional clusters and applies a Sync-Async synchronization modality: sites synchronize frequently with a nearby cluster aggregator over fast local links (synchronous cluster tier), while cluster aggregators push pre-aggregated updates asynchronously to a global aggregator over the WAN (asynchronous global tier). \sysname forms clusters based on both network proximity and power availability, and implements an adaptive synchronization policy which further reduces communication energy by adapting how often clusters synchronize to training progress. This structure shortens the synchronization barrier that idles GPUs and replaces per-site WAN transmissions each round with fewer, pre-aggregated transmissions at a lower frequency, reducing long-haul traffic. We evaluate \sysname at 100-site scale in a Flower-based simulation environment. \sysname matches or slightly improves the time-to-accuracy of single-tier baselines while reducing energy consumption by up to $3.9\times$.
\end{abstract}

\maketitle

\section{Introduction}
\label{sec:intro}
\begin{figure}[t]
    \centering
    \includegraphics[width=\columnwidth]{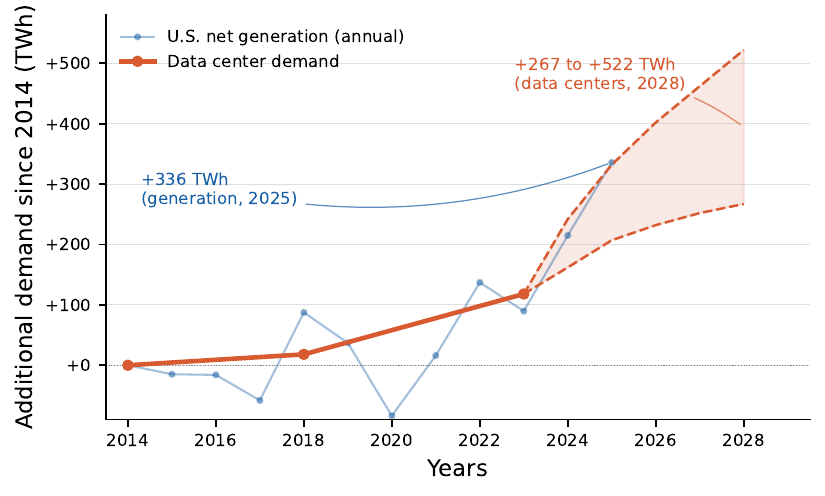}
    \caption{Growth in U.S. data center electricity demand against the growth in
    total grid generation~\cite{shehabi2024, eia_browser}.}
    \label{fig:demand_gap}
\end{figure}

The computational and power demands of large-scale AI training increasingly exceed what any single data center can supply~\cite{avelar2023ai, powertrip, goldmansachs2024}. U.S.\ data centers consumed 176~TWh of electricity in 2023, up from 76~TWh in 2018, an increase of 100~TWh in five years which is projected to grow by an additional 149--404~TWh by 2028~\cite{shehabi2024}. Over the same 2018--2023 window, total U.S.\ electricity generation grew by only 2.3~TWh~\cite{eia_browser}. \autoref{fig:demand_gap} shows the resulting divergence: AI-driven energy demand is growing by an order of magnitude faster than the grid that must supply it.

The shortfall extends beyond raw generation capacity; it is fundamentally a deliverability bottleneck. Even when new capacity is provided, grid interconnection constraints prevent immediate deployment. Currently, over 2{,}300\,GW of generation and storage remains stalled in transmission interconnection queues~\cite{lbnl_queued}. Furthermore, the median project now waits roughly five years from request to operation, up from under two years a decade ago~\cite{norris2025}. Consequently, new power cannot be provisioned on the timelines or at the locations required by AI training workloads, leaving individual datacenters strictly power-constrained.

When faced with strict power constraints at individual sites, operators often throttle workloads via power capping~\cite{sakalkar2020} or demand response~\cite{norris2025}. However, this approach idles provisioned GPUs and prolongs training times. Aggregating the available capacity across \emph{multiple} power-constrained sites mitigates this underutilization. Industry architectures are increasingly adopting this paradigm: Microsoft's ``AI superfactory'' interconnects datacenters across different U.S. states via dedicated wide-area networks (WANs) to train single models as a unified virtual cluster~\cite{guthrie2025}, while Google distributes the training of its largest foundation models across multiple campuses and metropolitan areas~\cite{google2026diloco}. Consequently, geo-distributed training across power-constrained sites is emerging as standard practice for large-scale AI workloads~\cite{sakalkar2020}.

However, geo-distribution introduces substantial communication overhead. Depending on site placement, synchronization rounds may cross regional or long-haul WAN 
links whose latency and bandwidth are much worse than within a single data center. This communication penalty can severely degrade, or entirely negate, the computational benefit provided by aggregate power. Consequently, realizing a net performance benefit from geo-distributed training is non-trivial. Prior work has demonstrated its viability: by strategically selecting the number and location of participating sites, it navigates the fundamental trade-off between distributed power capacity and communication delay~\cite{powertrip}.

Prior work, however, evaluates geo-distributed training almost entirely through the lens
of \emph{performance}: it optimizes time-to-accuracy and treats site power as a capacity
constraint, without directly accounting for the total \emph{energy} consumed to reach convergence.
Our central insight is that this performance benefit is obtained at a large and largely
avoidable energy cost because prior work largely relies on single-tier aggregation, where every site is exchanging model updates directly with a
central aggregator over the wide-area network (WAN) each synchronization round. 
Single-tier, all-to-one aggregation is fundamentally energy-inefficient due to three compounding factors. First, synchronization barriers force faster sites into an
idle state in which their GPUs continue to draw roughly 25--35\% of peak
power~\cite{Kandiah} while waiting for the most distant participant, so each round's
energy overhead scales with the latency of the slowest site in the pool of training sites. Second, every
site transmits full gradient updates directly over WAN links; as the site count grows, this redundant long-haul traffic dominates the energy
budget, and beyond a critical threshold adding more participants actively \emph{degrades}
energy efficiency because the communication penalty outweighs the computational gain. Third, a static synchronization frequency incurs a constant communication penalty even during later stages of training where updates yield diminishing convergence benefits~\cite{agarwal2021accordion}.

Building on this insight, \sysname exploits a structural property of geo-distributed
networks: intra-regional links exhibit round-trip latencies one to two orders of
magnitude lower than inter-continental links. \sysname is a hierarchical aggregation system
that organizes sites into regional clusters mirroring this physical network
structure and optimizes energy through three \emph{energy-efficiency policies}. 
First, a Sync-Async synchronization modality applies synchronous aggregation within each cluster to keep cluster models coherent at low cost, while asynchronous aggregation across clusters reduces the WAN-scale waiting barrier that would otherwise idle entire clusters. Second, power-aware agglomerative clustering groups sites on a joint criterion fusing network proximity with available power, producing clusters whose member sites have both low latency and high compute throughput. Third, an adaptive synchronization frequency assigns each cluster a global synchronization rate proportional to its communication cost and adaptively tapers WAN exchanges as training matures and parameter update magnitudes decrease. Together, these policies confine the majority of communication to fast regional links and reduce costly global exchange as the model converges, substantially lowering both idle and transmission energy while modestly improving time-to-accuracy.

Realizing these policies introduces significant challenges: \sysname must cluster
sites according to their power capacities and network proximity, determine the appropriate number and composition of clusters, modulate global synchronization frequency per-cluster and over time, and balance
synchronous against asynchronous inter-cluster aggregation without introducing model drift. Prior hierarchical federated learning explores similar structures but targets mobile edge computing, optimizing for
privacy and battery life under relatively static conditions~\cite{luo2025deep}; it does
not address the heterogeneous power capacities, WAN-bound communication, and energy objectives of large-scale data center training.

We hypothesize that a hierarchical aggregation structure mirroring the physical network, grouping geo-distributed sites into regional clusters with appropriate synchronization policies, can substantially reduce the energy cost of distributed training relative to a single-tier, all-to-one aggregation, while maintaining comparable time-to-accuracy. In evaluating this hypothesis, we make the following contributions:

\noindent
\textbf{Quantify energy inefficiency of single-tier aggregation.}
We show that single-tier geo-distributed training optimized for time-to-accuracy is highly energy-inefficient in three ways: idle waiting at synchronization barriers, excessive WAN transmissions, and convergence-agnostic
synchronization frequency.

\noindent
\textbf{Energy-efficient policies for geo-distributed training.}
We design \sysname, which introduces three energy-efficiency policies: (1)~a Sync-Async communication modality that eliminates WAN-scale waiting barriers while preserving intra-cluster model coherence (2)~agglomerative clustering on a joint network-proximity and power-availability criterion and (3)~an adaptive synchronization frequency that assigns per-cluster rates and tapers global WAN exchanges over training. 

\noindent
\textbf{Implementation and evaluation.}
We evaluate \sysname using a Flower-based~\cite{flower} simulation at 100-site scale. Our policies reduce the energy-to-accuracy by $3.9\times$ relative to single-tier baselines while maintaining or slightly
improving time-to-accuracy across multiple ML workloads.

\section{Background} 
\label{sec:background}

In this section, we first discuss the power constraints that make geo-distributed training
necessary (\autoref{sec:power_constraints}), then introduce geo-distributed model training and
the single-tier aggregation used by prior work (\autoref{sec:geo_distributed}). We then define the energy-inefficiency problem with the single-tier aggregation (\autoref{sec:single_tier}). 

\subsection{Power Constraints at a Single Site}
\label{sec:power_constraints}

A data center's computational throughput depends on several factors, including its hardware
resources, cooling capacity, and network bandwidth. Increasingly, the fundamental
constraint is power, because the electricity a site can draw is capped by its grid connection.
This constraint limits how many accelerators a site can run, regardless of how much hardware is physically installed~\cite{norris2025, powertrip}.

To use data center capacity efficiently, operators often oversubscribe their power budgets,
provisioning more hardware than the site could run at full power. As a result, a training job
must share the available power with co-located workloads, which compete for the same
grid allocation~\cite{sakalkar2020}. The power available for training at site $k$ is therefore
the time-varying residual $P_{\text{cap},k} - P_{\text{utilized},k}(t)$, where
$P_{\text{cap},k}$ is the site's rated power capacity and $P_{\text{utilized},k}(t)$ is the
power consumed by other workloads at time $t$. The residual power fluctuates 

and differs substantially across sites~\cite{sakalkar2020}.

When the residual power of a site drops, it can sustain fewer active accelerators, reducing its
training throughput proportionally. To avoid exceeding the available power, the workload is
throttled, either by capping power consumption or shedding load through demand response, which idles
provisioned hardware and delays training. This power shortage is not isolated to a single 
site. With power demand increasingly outpacing supply, no single site can reliably provide the sustained power that
large-scale model training demands. Moreover, new grid capacity cannot be brought online quickly
enough to close the gap, as interconnection queues now impose multi-year delays~\cite{lbnl_queued}. Pooling many such sites, however, aggregates far more power and compute
capacity than any single site can provide. This motivates the shift to geo-distributed training.

\subsection{Geo-Distributed Model Training}
\label{sec:geo_distributed}

Geo-distributed model training refers to the practice of training a single machine learning
model across multiple data centers that are geographically dispersed, potentially spanning
different cities, states, or continents. Traditional distributed training assumes co-located
nodes connected by high-bandwidth, low-latency interconnects, such as NVLink within a server and
InfiniBand across servers within a cluster~\cite{li2014parameterserver}. Geo-distributed training instead
operates over wide-area networks. These networks have high propagation delays (tens to hundreds
of milliseconds), limited and variable bandwidth, and heterogeneous compute across sites.

Existing systems geo-distribute training using two primary paradigms. Industry deployments, such as Microsoft's AI superfactory~\cite{guthrie2025}, partition the \emph{model} across sites. This approach extends pipeline or tensor parallelism over dedicated inter-site links to pool massive hardware arrays. However, model parallelism requires constant, high-bandwidth communication between sites. It also typically necessitates transporting massive volumes of training data across the network to the compute sites. These requirements make pipeline parallelism highly inefficient and brittle over a standard wide-area network (WAN). 

In contrast, Federated Learning (FL) partitions the \emph{data} across sites~\cite{mcmahan2017fedavg}. FL keeps data locally situated primarily to preserve user privacy. Our setting adopts this data-parallel architecture, but for an entirely different reason. In power-constrained environments, data parallelism is structurally better than model parallelism. Because each site computes independently on its local data partition and only synchronizes model parameters periodically, this decoupled approach tolerates high WAN latencies and power fluctuations. 

Following this paradigm, we assume the training dataset is already distributed and locally available at the participating sites. While this mirrors the mechanical structure of FL, we do not partition data for privacy. Instead, sites participate because they possess spare power capacity (\autoref{sec:power_constraints}).

Under data parallelism, the synchronization mechanism, namely how often, with whom, and over
what network path model updates are exchanged, is the main challenge, as the WAN introduces
communication costs that are orders of magnitude higher than those in co-located
settings~\cite{wan2024netstorm}. Existing systems organize this synchronization as \emph{single-tier
aggregation}, where every participating site exchanges its model update directly with one
central aggregator over the WAN each round~\cite{li2014parameterserver}. A round proceeds
synchronously. The aggregator broadcasts the current global model, each site trains locally for
a fixed number of steps, and the aggregator averages the returned updates into a new global
model. The time to transfer a model update of size $D_m$ between a site $k$ and the aggregator
is
\begin{equation}
T_{k,\text{comm}} = \frac{2 d_k}{c} + \frac{D_m}{B_k},
\label{eq:comm_delay}
\end{equation}
where $d_k$ is the physical distance between them, $c$ is the propagation speed of the signal,
and $B_k$ is the available bandwidth on the link~\cite{powertrip}. The first term is
the round-trip propagation delay; the second is the transmission time. A round cannot complete
until every site has returned its update, so the round's communication time is set by the
slowest participant, $T_{\text{comm}} = \max_{k \in S} T_{k,\text{comm}}$.
 
Prior systems address this communication cost by tuning which and how many sites
participate~\cite{powertrip}. Others reduce the data exchanged per round, through
gradient compression~\cite{lin2018deepgradient} or additional local steps between
synchronizations~\cite{douillard2023diloco}. In all cases, however, the structure remains
single-tier: every site synchronizes directly with a single global aggregator.

\subsection{Inefficiency of Single-Tier Aggregation}
\label{sec:single_tier}

Single-tier aggregation techniques are designed to maximize
throughput and minimize time-to-convergence, and are commonly
deployed in edge or mobile environments where transferred model
updates are small (kilobytes to
megabytes)~\cite{mcmahan2017fedavg} and wide-area bandwidth
limitations remain manageable. In geo-distributed datacenter
training, however, model updates are massive, often gigabytes in
size~\cite{douillard2023diloco}. At this scale, WAN communication
becomes a major bottleneck, degrading both performance and energy
efficiency. Prior work primarily optimizes for training time and
overlooks the energy overhead of long-haul communication and idle
resources.

\begin{figure}[t]
    \centering
    \includegraphics[width=\columnwidth]{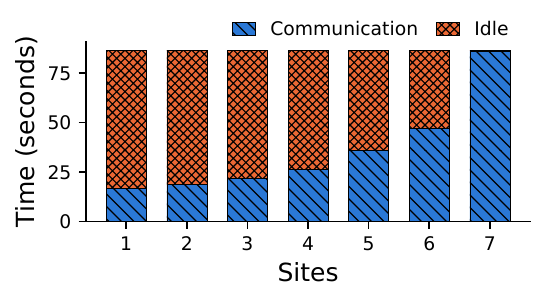}
    \caption{Per-site communication and idle time in one synchronous round, ordered by increasing distance from the aggregator.}
    \label{fig:straggler_idle}
\end{figure}

To formalize this overhead, we consider the energy cost of a
single training round. Each site passes through three activity
states: computing, communicating, and idling at the
synchronization barrier while waiting for all other participants
to report. The accelerators remain powered in each state but draw
power at different rates. Letting $t_{\text{comp},k}$,
$t_{\text{comm},k}$, and $t_{\text{idle},k}$ denote the time site
$k$ spends in each state, its round energy is

\begin{equation}
E_k = P_{\text{frac},k} \left( t_{\text{comp},k} + \alpha\, t_{\text{comm},k} + \beta\, t_{\text{idle},k} \right),
\label{eq:energy_model}
\end{equation}
where $P_{\text{frac},k}$ is the normalized power allocation at
site $k$, proportional to the power it draws while computing
(\autoref{sec:power_constraints}), and $\alpha, \beta \in (0,1)$
scale this draw during communication and idling, since
accelerators consume less power when not actively computing.
Summed over all sites and aggregators, the total decomposes into
$E = E_{\text{compute}} + E_{\text{comm}} + E_{\text{idle}}$.

This decomposition reveals two properties that shape our design.
First, compute energy is essentially fixed: it reflects
the work needed to reach a target accuracy, which reorganizing
sites does not change. Power throttling
(\autoref{sec:power_constraints}) stretches $t_{\text{comp},k}$
when $P_{\text{frac},k}$ decreases, but their product remains
constant. Second, because every term is power multiplied by time,
reducing the duration of communication or idling reduces the
corresponding energy directly. The reducible overhead is therefore
the communication and idle energy, and single-tier aggregation
inflates exactly these two components in three ways.

First, it inflates idle energy (the
$\beta\,P_{\text{frac}}\,t_{\text{idle}}$ term of
Eq.~\ref{eq:energy_model}). Each synchronous round cannot finish
until the most distant site reports, so faster sites sit at the
barrier with their accelerators powered but idle. This energy overhead
grows with the latency spread across sites, which is large
over the WAN. \autoref{fig:straggler_idle} illustrates this directly. Sites near the
aggregator finish their transfers well before the most distant site, yet
none can proceed until the slowest participant arrives. In a single-tier
design, this waiting repeats every round, paying for accelerator power with no training progress.

Second, it inflates communication energy ($\alpha\,P_{\text{frac}}\,t_{\text{comm}}$). Every site
sends its full update over the WAN each round. Because
nearby sites transmit independently rather than combining their
updates regionally first, many of these transmissions are
redundant: the same information could reach the aggregator in a
single pre-aggregated message per region. This redundant traffic
grows with site count, and beyond a critical number, adding sites
degrades energy efficiency even as it continues to improve
time-to-accuracy.

Third, communication energy is spent even when it yields
little benefit. A fixed synchronization frequency has the same
per-round WAN cost throughout training, including late stages
where gradients are small and synchronization barely
improves accuracy.

Existing single-tier techniques do not address these overheads.
Backup workers~\cite{chen2016revisiting} mitigate stragglers but
require spare compute capacity that power-constrained sites lack.
Collective operations like ring
all-reduce~\cite{sergeev2018horovod} assume uniform,
high-bandwidth interconnects rather than slow, heterogeneous WAN
links. Asynchronous methods remove the synchronization barrier
but still incur a full WAN round-trip per step and introduce
model drift. Recent work optimizes site
selection~\cite{powertrip} or synchronization
frequency~\cite{douillard2023diloco}, reducing some overhead, but
every site still reports directly to a global aggregator,
leaving these energy inefficiencies unresolved.

\vspace{-0.3cm}

\section{\sysname Design}
\label{sec:design}
\begin{figure*}[t]
\centering
\resizebox{\textwidth}{!}{
\begin{tikzpicture}[
    >=latex,
    ga/.style     ={draw,fill=blue!12,rounded corners,thick,minimum width=3.0cm,minimum height=0.75cm,font=\footnotesize\bfseries},
    ra/.style     ={draw,fill=violet!12,rounded corners,thick,minimum width=2.2cm,minimum height=0.7cm,align=center,font=\scriptsize},
    ca/.style     ={draw,fill=green!12,rounded corners,thick,minimum width=1.1cm,minimum height=0.55cm,font=\scriptsize},
    site/.style   ={draw,rounded corners=2pt,minimum width=0.6cm,minimum height=0.4cm,line width=0.5pt},
    module/.style ={draw,fill=gray!10,rounded corners,thick,align=center,font=\scriptsize,text width=2.4cm,inner sep=3pt},
    inner/.style  ={<->,thick,green!55!black},
    outer/.style  ={<->,thick,dashed,red!70!black},
    modlink/.style={->,thick,dotted,gray},
  ]

  \draw[rounded corners=5pt,fill=blue!4,draw=blue!30,dashed,line width=0.6pt]
        (-6.0, -0.3) rectangle (-4.0, 1.6);

  \foreach \x/\s/\n in {
      -5.6/50/a1, -5.0/20/a2, -4.4/42/a3,
      -3.1/55/b1, -2.5/25/b2, -1.9/35/b3,
      -0.6/35/c1,  0.0/50/c2,  0.6/22/c3,
       1.9/45/d1,  2.5/30/d2,  3.1/45/d3,
       4.4/10/e1,  5.0/60/e2,  5.6/30/e3}
    \node[site,fill=orange!\s] (\n) at (\x,0) {};

  \node[ca] (CA1) at (-5.0, 1.2) {CA$_1$};
  \node[ca] (CA2) at (-2.5, 1.2) {CA$_2$};
  \node[ca] (CA3) at ( 0.0, 1.2) {CA$_3$};
  \node[ca] (CA4) at ( 2.5, 1.2) {CA$_4$};
  \node[ca] (CA5) at ( 5.0, 1.2) {CA$_5$};

  \node[ra] (RA1) at ( 1.25, 2.4) {Regional\\[-0.5ex]Aggregator 1};
  \node[ra] (RA2) at ( 5.0,  2.4) {Regional\\[-0.5ex]Aggregator 2};

  \node[ga] (GA)  at (-1.0, 3.6) {Global Aggregator};

  \foreach \n in {a1,a2,a3} \draw[inner] (\n.north) -- (CA1.south);
  \foreach \n in {b1,b2,b3} \draw[inner] (\n.north) -- (CA2.south);
  \foreach \n in {c1,c2,c3} \draw[inner] (\n.north) -- (CA3.south);
  \foreach \n in {d1,d2,d3} \draw[inner] (\n.north) -- (CA4.south);
  \foreach \n in {e1,e2,e3} \draw[inner] (\n.north) -- (CA5.south);

  \draw[outer] (CA1.north) -- (GA.south west);
  \draw[outer] (CA2.north) -- (GA.south);

  \draw[outer] (CA3.north) -- (RA1.south west);
  \draw[outer] (CA4.north) -- (RA1.south east);
  \draw[outer] (CA5.north) -- (RA2.south);

  \draw[outer] (RA1.north) -- (GA.south east);
  \draw[outer] (RA2.north west) -- (GA.east);

  \node[module] (CL)  at (-6.4, 3.4)
    {\textbf{Clustering Engine}\\[1pt] \tiny power $+$ network\\ \tiny auto-$K$ \& dynamic};
  \node[module] (SCH) at ( 6.6, 3.4)
    {\textbf{Agg. Scheduler}\\[1pt] \tiny inner sync, outer async\\ \tiny adaptive per-cluster $L_c$};

  \draw[modlink] (CL.south)  -- (-6.4, 1.2) -- (CA1.west);
  \draw[modlink] (SCH.south) -- ( 6.6, 1.2) -- (CA5.east);

  \node[font=\scriptsize,gray] at (-4.2, 2.4) {$H{=}2$};
  \node[font=\scriptsize,gray] at ( 2.6, 3.35) {$H{=}3$};

  \draw[inner] (-5.3,-0.9) -- (-4.8,-0.9);
  \node[right,font=\scriptsize] at (-4.7,-0.9) {inner: sync (local)};
  
  \draw[outer] (-2.2,-0.9) -- (-1.7,-0.9);
  \node[right,font=\scriptsize] at (-1.6,-0.9) {outer: async (WAN)};
  
  \node[site,fill=orange!40,minimum width=0.4cm,minimum height=0.25cm] at (0.8,-0.9) {};
  \node[right,font=\scriptsize] at (1.0,-0.9) {shading $\propto$ available power};
  
  \draw[rounded corners=2pt,fill=blue!4,draw=blue!30,dashed] (4.1,-1.05) rectangle (4.5,-0.75);
  \node[right,font=\scriptsize] at (4.6,-0.9) {cluster boundary};

\end{tikzpicture}
}
\caption{\sysname's architecture. The left side depicts the default two-tier ($H{=}2$) topology; the right illustrates a three-tier ($H{=}3$) extension using regional aggregators.}
\label{fig:system_overview}
\end{figure*}
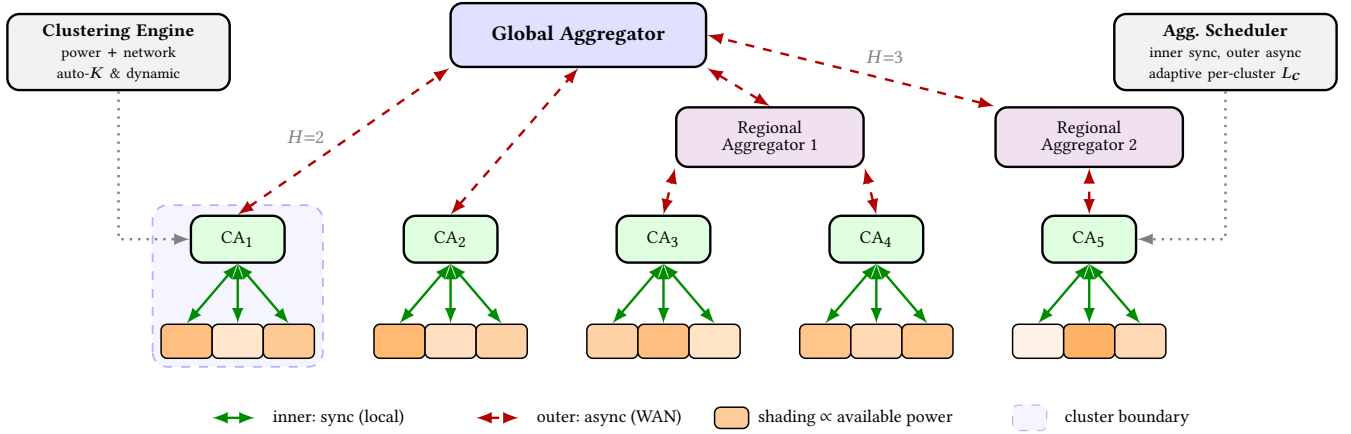

To address the energy inefficiencies of single-tier geo-distributed training in power-constrained data centers, we present \sysname. \sysname utilizes hierarchical aggregation, exploiting the spatial heterogeneity of wide-area networks: links between data center sites in the same region offer bandwidth and latency an order of magnitude better than inter-continental ones~\cite{wan2024netstorm}. By confining most synchronization to these fast regional links, hierarchical aggregation reduces communication energy, because each cluster forwards a single pre-aggregated update over the WAN in place of many individual ones. It also reduces idle energy, because sites wait only for others in their own cluster rather than for all participants.

Hierarchical aggregation is not a new concept. Hierarchical Federated Learning (HFL)~\cite{liu2020clientedgecloud, luo2025deep, zhang2024communication} applies this principle in mobile and edge computing, where it relieves communication bottlenecks at the central server and accommodates resource-constrained edge devices. \sysname borrows the same structural intuition but applies it in a different context and with a different objective: minimizing energy inefficiencies under strict power constraints. Unlike HFL, where edge devices can sleep when idle and consume negligible power, datacenter accelerators draw substantial power even when stalled at a synchronization barrier, making every wait an energy cost rather than a delay. Moreover, each site in our setting contributes power, and reaching more distant sites to pool additional power incurs proportionally higher communication energy, a trade-off between power and communication cost that does not arise in the edge setting.

Implementing this hierarchical structure for energy-efficient geo-distributed training raises three technical challenges, each corresponding to one of the energy overheads identified in \autoref{sec:single_tier}. First, the hierarchy introduces synchronization at two levels, within clusters and between clusters, and how each level synchronizes directly determines idle energy. If all clusters must wait for the slowest one before the global model is updated, the energy overhead due to straggler idling of single-tier aggregation reappears at the cluster level (\autoref{sec:sync_modality}). Second, how sites are grouped into clusters determines both communication and idle energy simultaneously. A poor grouping can place nearby sites in different clusters, reintroducing redundant long-haul traffic, or strand a low-power site in a fast cluster, creating a straggler that idles every other member (\autoref{sec:dynamic_clustering}). Third, how often clusters synchronize presents a trade-off: more frequent synchronization incurs higher WAN communication energy, while less frequent synchronization leads to model drift and slower convergence (\autoref{sec:agg_frequency}).

\sysname addresses these challenges through an energy-efficiency policy that resolves each one dynamically at runtime, adapting to each site's available power and the model's convergence progress rather than a fixed configuration set before training begins. \autoref{fig:system_overview} shows the resulting architecture. We discuss each component in detail in the following subsections, beginning with the synchronization modality that forms the foundation of the system.

\subsection{Synchronization Modality}
\label{sec:sync_modality}

Given $N$ sites, \sysname organizes them into $K$ regional clusters, each managed by a \emph{cluster aggregator} (CA), with the CAs reporting to a single \emph{global aggregator} (GA). Model training uses a two-tier update strategy: frequent updates occur at the \emph{cluster level (within each cluster)}, while infrequent updates occur at the \emph{global level (across all clusters)}.

In \autoref{fig:system_overview}, these are the green intra-cluster links and the red inter-cluster links, respectively.

The first design decision is how each tier synchronizes. Each tier can operate either synchronously, where all participants must report before aggregation proceeds, or asynchronously, where updates are merged as they arrive. 
This yields four configurations.
\begin{itemize}[leftmargin=*, topsep=2pt, itemsep=2pt, parsep=0pt]
    \item \textbf{Sync-Sync.} Both tiers are synchronous. Models stay fully coherent, but faster sites and faster clusters must wait for the slowest one at both levels.

    \item \textbf{Sync-Async.} The cluster tier is synchronous; the global tier is asynchronous. Clusters stay coherent internally, while they no longer wait for each other at the global tier. 

    \item \textbf{Async-Sync.} The cluster tier is asynchronous; the global tier is synchronous. Sites within a cluster can finish and be aggregated at different times, but clusters still wait for others to send updates.

    \item \textbf{Async-Async.} Both tiers are asynchronous. No participant waits for any other, but updates can be aggregated using stale information at both levels, which slows convergence.
\end{itemize}

At each tier, the synchronization mode is chosen based on how much participants differ in their completion times.

\noindent
\textbf{Cluster tier (intra-cluster).} In each round, the CA broadcasts its current model, sites train locally on their data partitions, and the CA averages the returned updates. \sysname aggregates the model updates at the cluster tier \emph{synchronously}. This is because the sites within a cluster are geographically proximate and share similar network characteristics, so their completion times are close to each other. The wait imposed on faster sites is therefore short, and the idle energy accumulated during that wait is small. Synchronous aggregation also keeps the cluster model coherent. Running the cluster tier asynchronously can reduce the wait further, but since completion times are similar, the marginal energy savings would come at the cost of model coherence. 

\noindent
\textbf{Global tier (inter-cluster).} At the global tier, the GA aggregates the model updates from clusters \emph{asynchronously}. Clusters connect to the global aggregator over the WAN, and their completion times differ significantly due to differences in link latency and aggregate compute. Waiting for the slowest cluster each round would reintroduce idle energy overhead at the cluster level, negating the benefit of the hierarchy. Instead, each CA pushes its update to the GA as soon as it finishes, without waiting for other clusters. The GA merges arriving updates into the global model, weighting each by the contributing cluster's size. To handle updates arriving at different times, the GA maintains a buffer of size $B$. Each time $B$ cluster updates have accumulated, the GA aggregates them into the global model, weighting each update inversely by its staleness, the number of global rounds elapsed since it was computed. This ensures that more recent updates have greater influence on the global model, while older updates are not discarded but down-weighted. Because the hierarchy replaces $N$ individual WAN transmissions per round with only $K$ cluster-level transmissions every few rounds, the total long-haul communication drops substantially, directly reducing communication energy.

\sysname therefore applies the \textbf{Sync-Async} configuration: synchronous at the cluster tier and asynchronous at the global tier. This confines waiting to the cluster tier, where the cost is low, and removes it from the global tier, where the cost is high. We evaluate all four configurations in \autoref{sec:eval_sync}. We integrate this Sync-Async modality as the core communication loop of our runtime policy (\autoref{alg:powerscale}, Lines 10--18).

\subsection{Clustering Policy}
\label{sec:dynamic_clustering}

With the synchronization modality established, the second component of \sysname's policy determines how sites are grouped into clusters. The grouping determines both communication and idle energy: placing nearby sites in the same cluster confines most traffic to fast regional links, while balancing power across clusters prevents any one cluster from being slowed by a low-power site. \sysname's clustering policy addresses three decisions: the clustering algorithm, the criterion on which sites are grouped, and how the grouping is maintained as conditions change during training.

\noindent
\textbf{Clustering Algorithm.}
\sysname supports two clustering algorithms, selected based on the deployment scenario.

\textbf{K-means}~\cite{kmeans} partitions sites along the network-distance axis, minimizing within-cluster variance in distance to the global aggregator. It is preferred when the cluster count is known in advance, for example when it corresponds to fixed geographic regions, and when low computational overhead during re-clustering is a priority.

\textbf{Agglomerative clustering}~\cite{agglomerative} is \sysname's default for deployments where the cluster count is not fixed or where site conditions change frequently. It operates on a combined feature that fuses network distance and power availability,
\begin{equation}
f_k = w_d\,\tilde{d}_k + w_p\,(1 - \tilde{P}_{\text{frac},k}),
\label{eq:cluster_feature}
\end{equation}
where $\tilde{d}_k$ and $\tilde{P}_{\text{frac},k}$ are the normalized distance and power fraction of site $k$, and $w_d$ and $w_p$ are weights that control the relative importance of the two features (we use $w_d{=}0.7$, $w_p{=}0.3$). Starting from singleton clusters, the algorithm repeatedly merges the closest pair of clusters under $f_k$ until the target count is reached. The merge sequence produces a dendrogram, a hierarchical tree of clusters that enables efficient re-clustering at runtime: cutting the tree at a different height yields a new grouping without re-solving from scratch. The nested structure naturally supports multi-level hierarchies.

\noindent
\textbf{Clustering Criterion.}
\sysname groups sites jointly on two properties: network distance and available power. Network distance $d_k$ determines the communication latency between a site and its cluster aggregator, as given by the delay model of Eq.~\ref{eq:comm_delay}. Available power fraction $P_{\text{frac},k} \in (0,1]$ determines a site's available compute throughput (\autoref{sec:power_constraints}). \sysname combines these into a score that reflects the compute a site contributes relative to the communication cost it introduces,
\begin{equation}
\text{Score}(k) = \frac{P_{\text{frac},k}}{T_{k,\text{comm}}},
\label{eq:site_score}
\end{equation}
where $T_{k,\text{comm}}$ is the transfer time from site $k$ to the global aggregator for a reference update size (Eq.~\ref{eq:comm_delay}). This score is used to rank sites when determining active participation (\autoref{sec:cluster_count}).

The two properties serve complementary roles. Grouping sites by proximity keeps intra-cluster transfer times short, directly reducing communication energy. Balancing power across clusters ensures that no cluster is consistently paced by a low-power site, reducing idle energy at the synchronization point. Clustering on network distance alone can strand a low-power site inside an otherwise fast cluster, raising idle energy. Clustering on power alone can group distant sites together, inflating communication cost. 

\noindent
\textbf{Cluster Count and Size.}
\label{sec:cluster_count}
Deciding the number of clusters $K$ involves a trade-off. A small $K$ produces large clusters that amortize global synchronization across many sites, but large clusters increase the probability that at least one site finishes significantly later than the others, extending the synchronization wait. A large $K$ produces smaller, more homogeneous clusters, but increases the number of cluster aggregators that must transmit over the WAN each global round, raising communication energy.

\sysname selects $K$ automatically by sweeping candidate values up to $\min(2\sqrt{N}, N/2)$ and estimating per-round cost for each under the delay model of Eq.~\ref{eq:comm_delay}. $K$ with the lowest estimated cost is selected, favoring larger values on near-ties for greater parallelism.

Individual cluster sizes vary because \sysname groups sites by feature similarity (Eq.~\ref{eq:cluster_feature}) rather than enforcing equal counts. To prevent any cluster from growing to a size that slow completions become likely, \sysname bounds each cluster's size around the mean. Assuming independent completion events across sites, the probability that at least one site in a cluster of size $N_c$ causes a delay is
\begin{equation}
P(\text{delay}) = 1 - \big(1 - p_s\big)^{N_c},
\label{eq:straggler_prob}
\end{equation}
where $p_s$ is the per-site probability of a slow completion in a given round. This probability grows with $N_c$, so \sysname limits cluster sizes to keep it below an acceptable threshold.

\noindent
\textbf{Dynamic vs. Static Clustering.}
A \emph{static clustering} derived from conditions at the start of training is optimal only for those initial conditions. Available power varies over time as co-located production workloads compete for grid capacity (\autoref{sec:power_constraints}). As power budgets shift, a grouping that was well-balanced initially may place power-reduced sites into clusters where they extend the synchronization wait, increasing idle energy, or may exclude high-powered sites that have become available, leaving compute capacity unused.

\sysname therefore presents a \emph{dynamic clustering} technique which re-evaluates the clustering during training, triggered periodically and whenever a significant change in site power is detected. Using the existing dendrogram, \sysname updates the current grouping incrementally, splitting clusters that have grown too large and merging those that have become too small, re-solving from scratch only when incremental updates are insufficient. This approach keeps the grouping aligned with current conditions while minimizing re-clustering overhead. The runtime policy invokes these clustering mechanisms natively during initialization and continuously monitors for necessary dynamic re-assignments (\autoref{alg:powerscale}, Lines 2--5 and 19--22).

\subsection{Synchronization Frequency}
\label{sec:agg_frequency}

The third component of \sysname's policy controls how often each cluster synchronizes with the global aggregator. This is governed by the local round ratio $L$, the number of intra-cluster rounds between consecutive global-tier aggregations. 
A larger $L$ means more local rounds between global synchronizations, which reduces the frequency of WAN transfers: over $R$ total training rounds, the system performs only $R/L$ global exchanges, so increasing $L$ directly reduces total communication energy. However, fewer global synchronizations allow cluster models to diverge from the global average, potentially requiring additional rounds to reach target accuracy. The design challenge is choosing $L$ large enough to meaningfully reduce communication energy without increasing total training rounds to the point where the additional compute energy offsets the savings. \sysname supports two frequency policies.

\noindent
\textbf{Fixed frequency.} The simplest policy assigns a single fixed $L$
to all clusters and holds it constant throughout training. Every cluster
synchronizes at the same interval regardless of its communication cost or
the model's convergence state. This policy cannot adapt to either spatial
or temporal variation and serves as the lower-bound baseline.

\noindent
\textbf{Uniform frequency.} The uniform policy also applies the same $L$
to all clusters within each round, but allows $L$ to change over time as
training progresses. This captures temporal diminishing returns, reducing
global synchronization as updates become incremental, but it cannot
account for heterogeneous communication costs across clusters: a cluster
with high WAN cost synchronizes at the same frequency as one with low
WAN cost.

\noindent
\textbf{Adaptive frequency.}
Clusters vary in their distance to the global aggregator, and the marginal value of each global synchronization decreases as training progresses. \sysname's adaptive frequency policy accounts for both spatial heterogeneity across clusters and temporal diminishing returns over the course of training.

It first assigns each cluster a per-cluster baseline ratio proportional to its relative communication cost:
\begin{equation}
L_c = \min\!\left( L_{\max},\; \max\!\left(1,\;
  \left\lfloor L_{\text{base}} \cdot
  \frac{\bar{T}^{\text{global}}_{c}}{\bar{T}^{\text{local}}_{c}}
  \right\rfloor \right)\right),
\label{eq:asymmetric_L}
\end{equation}
where $L_{\text{base}}$ is a configurable baseline ratio shared across all sites that sets the scale of all per-cluster values, $\bar{T}^{\text{global}}_{c}$ and $\bar{T}^{\text{local}}_{c}$ are cluster $c$'s mean inter- and intra-cluster transfer times (Eq.~\ref{eq:comm_delay}), and $L_{\max}$ caps the local rounds any cluster may run between global synchronizations, preventing indefinite model drift. Clusters with a high global-to-local cost ratio receive a larger $L_c$, synchronizing less often and spending less communication energy per round of training. 

The adaptive policy then modulates each $L_c$ over time based on observed training progress. As training matures, gradient magnitudes decrease and parameter updates become incremental~\cite{douillard2023diloco}; paying the same WAN energy in late training as in early training wastes energy on communication that barely improves accuracy. To avoid this, cluster $c$ defers global synchronization while its model is changing minimally, and triggers one when its accumulated local update grows large or its accuracy improvement stalls:
\begin{equation}
\big\| \theta_c^{(\ell)} - \theta_c^{(0)} \big\| > \delta
\quad\text{or}\quad
\Delta\alpha_c < \epsilon \;\text{for}\;
P_{\text{patience}} \;\text{consecutive rounds},
\label{eq:conv_sync}
\end{equation}
where $\theta_c^{(\ell)}$ is cluster $c$'s model after $\ell$ local rounds since its last global synchronization, and $\theta_c^{(0)}$ is the model it received from the global aggregator at the start of that interval. The first condition triggers a global synchronization when the cluster's local model has diverged significantly from the last globally aggregated model, ensuring that clusters undergoing substantial learning propagate their updates to the rest of the system before divergence becomes harmful. The second condition triggers a global synchronization when the cluster's local accuracy has not improved by more than $\epsilon$ for $P_{\text{patience}}$ consecutive rounds; in this case, the cluster has stalled locally and replaces its model with the current global average, which incorporates learning from all other clusters and may help the stalled cluster escape its plateau. Together, these conditions concentrate global communication in two regimes: early in training, when models change rapidly and the magnitude condition occurs frequently, and during local stalls, when a cluster needs external progress to continue improving. As training matures and updates become incremental, neither condition occurs often, and the interval between global rounds grows naturally, reducing communication energy. The cap $L_{\max}$ guarantees that every cluster synchronizes at least once within $L_{\max}$ local rounds regardless of these conditions, placing a hard bound on model drift.

When \sysname reassigns sites between clusters (\autoref{sec:dynamic_clustering}), the affected clusters' $\bar{T}^{\text{global}}_{c}$ and $\bar{T}^{\text{local}}_{c}$ change. After each reassignment, \sysname recomputes these values from the updated membership and recalculates $L_c$ via Eq.~\ref{eq:asymmetric_L} before resuming adaptive modulation. This ensures the frequency policy remains consistent with the current cluster structure. We evaluate both policies and their interaction with dynamic clustering in \autoref{sec:eval_frequency}. This adaptive frequency mechanism directly controls the condition for breaking the local training loop and triggering a global synchronization (\autoref{alg:powerscale}, Lines 6 and 11--16).

\subsection{Summary}
\label{sec:composed_policy}

\autoref{alg:powerscale} composes the three preceding mechanisms into a single runtime policy. The algorithm ties together initialization (clustering and $L_c$ assignment), the training loop (synchronous inner aggregation, adaptive outer synchronization triggers, and staleness-weighted global merging), and periodic reclustering as power conditions shift. 

Note that the policies outlined above assume a two-level hierarchy ($H{=}2$: sites $\rightarrow$ CAs $\rightarrow$ GA), placing the aggregation boundary at the WAN. Deeper hierarchies ($H > 2$) introduce additional pre-aggregation tiers that further consolidate WAN traffic, reducing per-round communication energy, but each additional tier accumulates model drift, which can increase the total rounds required to reach target accuracy. Whether the savings outweigh this cost depends on scale: at larger site counts, traffic reduction from an additional tier is more pronounced, while per-hop drift is bounded by $L_{\max}$. We adopt $H{=}2$ as the default and extend it to $H{=}3$ in \autoref{sec:eval_general} to characterize this trade-off. \sysname does not attempt to optimize depth. Determining the $H$ value that minimizes total energy for an arbitrary deployment is a hard joint optimization problem across network structure, power distributions, and convergence dynamics. This is out of scope and reserved for future work.

\begin{algorithm}[t]
\caption{\sysname Runtime Policy}
\label{alg:powerscale}
\begin{algorithmic}[1]
\Require $N$ sites with distances $d_{ij}$ and power fractions
  $P_{\text{frac},k}$; thresholds $L_{\max}$, $\tau_P$, $\delta$,
  $\epsilon$, $P_{\text{patience}}$; reclustering interval
  $R_{\text{recluster}}$
\Statex
\Statex \textbf{\emph{// Initialization
  (\autoref{sec:dynamic_clustering})}}
\State Rank sites by accuracy gain per unit energy
  (Eq.~\ref{eq:site_score}); select active set
\State Choose $K$ minimizing estimated per-round time; form
  clusters $\{\mathcal{C}_c\}$ using joint feature $f_k$
  (Eq.~\ref{eq:cluster_feature})
\State Designate site nearest each cluster centroid as cluster
  aggregator (CA); designate GA
\State Assign each cluster baseline $L_c$
  (Eq.~\ref{eq:asymmetric_L})
\Statex
\Statex \textbf{\emph{// Training loop}}
\For{each global round $t$}
  \For{each cluster $c$ \textbf{in parallel}}
    \State $\ell_c \gets 0$
    \Repeat
      \State CA broadcasts $\theta_c$; sites compute local
        updates; CA averages \Comment{sync inner}
      \State $\ell_c \gets \ell_c + 1$
    \Until{$\lVert \theta_c^{(\ell_c)} - \theta_c^{(0)} \rVert
      > \delta$ \textbf{or} $\Delta\alpha_c < \epsilon$ for
      $P_{\text{patience}}$ rounds \textbf{or}
      $\ell_c \geq L_{\max}$}
    \State CA pushes $\theta_c$ to GA with timestamp $t_c$
      \Comment{async outer}
  \EndFor
  \State GA incorporates each arriving $\theta_c$ into global
    model $\theta$, weighted by $1/(t - t_c + 1)$
  \Statex \textbf{\emph{// Reclustering check}}
  \If{$t \bmod R_{\text{recluster}} = 0$ \textbf{or}
    $\exists\, k: |P_{\text{frac},k}^{(t)} -
    P_{\text{frac},k}^{(\text{last})}| > \tau_P$}
    \State Migrate affected sites to nearest centroid;
      recompute $L_c$ for modified clusters
  \EndIf
\EndFor
\end{algorithmic}
\end{algorithm}

\section{Implementation}
\label{sec:implementation}
\sysname is implemented in
Python using PyTorch for model execution and Flower-compatible client
and strategy interfaces. An experiment configuration is first resolved into an explicit
topology containing leaf sites, aggregation nodes, and their parent--child relationships.
The runtime materializes this topology recursively: each leaf node wraps a data partition
and local training task, while each internal node presents the same \texttt{fit} and
\texttt{evaluate} interface as a client to its parent. This recursive representation lets
the same execution path instantiate a single-tier baseline, the default two-tier hierarchy, or a
deeper hierarchy without duplicating the training logic.

Each aggregation node maintains its current model, local-round counter, and synchronization
state. During a local step, it distributes a cloned model to its children, collects their
updates, and forms a sample-weighted aggregate. Leaf clients execute the actual PyTorch
training loop on their assigned partitions; a bounded worker pool runs independent sibling
clients concurrently while preserving the dependency order imposed by the hierarchy. At the
cluster tier, this procedure implements the synchronous inner loop. At the global tier, the
runtime can pass completed cluster updates to a buffered aggregator, which merges the buffer
using the configured staleness weights. The same aggregation primitives are also used by the
single-tier and fixed-policy configurations, ensuring that policies differ in topology and control
state rather than in the underlying model-update path.

The adaptive mechanisms are implemented as a controller separate from the training
primitives. At global-round boundaries, the controller maintains active-site membership,
cluster assignments, and the synchronization interval for each cluster. When a clustering
update changes the hierarchy, the runtime rebuilds the affected internal aggregation nodes
from the updated assignment while reusing the existing leaf clients and their data
partitions. It then propagates the revised synchronization intervals to the corresponding
nodes before the next round begins. A structured per-round log records model metrics,
elapsed execution time, participating sites, cluster members, synchronization intervals,
and asynchronous-aggregation metadata. This separation keeps the adaptive policy explicit
and makes every hierarchy and synchronization decision available to the analysis pipeline.

\section{Evaluation}
\label{sec:evaluation}
In this section, we present our evaluation methodology (\autoref{sec:eval_methodology}), validate the
energy inefficiency of single-tier aggregation (\autoref{sec:eval_inefficiency}), then
evaluate each component of \sysname's energy-efficiency policy in the order they appear in the
design: synchronization modality (\autoref{sec:eval_sync}), clustering policy
(\autoref{sec:eval_clustering}), and synchronization frequency
(\autoref{sec:eval_frequency}). We close with generalizability across datasets and the
hierarchy-depth trade-off (\autoref{sec:eval_general}).

\subsection{Evaluation Methodology}
\label{sec:eval_methodology}

We present our evaluation methodology, covering the experimental setup, training
datasets and models, baseline policies, evaluation metrics, and power and
communication profiling.

\noindent
\textbf{Experimental setup.} We evaluate \sysname in an emulated environment of
100 geo-distributed sites, following the methodology of prior work~\cite{powertrip}.
Each site is assigned a distinct geographic location, network profile, and power availability.
Because the number of emulated sites exceeds our physical GPUs, training tasks for multiple
virtual sites share a GPU and run sequentially, while we compute a simulated wall-clock time
that reflects execution under perfect parallelism. The communication time for each transfer
follows Eq.~\ref{eq:comm_delay}, and the computation time scales inversely with the site's
power fraction $P_{\text{frac},k}$ (\autoref{sec:power_constraints}). 

\noindent
\textbf{Datasets and models.} \autoref{tab:datasets} summarizes the datasets. Our primary
evaluation uses CIFAR-10~\cite{cifar10} with a ResNet-18~\cite{he2016resnet}, trained to a
target accuracy of 94\%. We demonstrate generalizability on the Shakespeare next-character
prediction corpus with a two-layer LSTM and on EMNIST image classification with a CNN
(\autoref{sec:eval_general}). The data are partitioned IID across sites, consistent with our system
model in which geo-distribution is driven by power rather than data locality
(\autoref{sec:design}).

\begin{table}[t]
\centering
\small
\begin{tabular}{@{}lllll@{}}
\toprule
\textbf{Dataset} & \textbf{Task} & \textbf{Model} & \textbf{Classes} & \textbf{Samples} \\
\midrule
CIFAR-10    & Image Class.    & ResNet-18 & 10 & 60{,}000 \\
Shakespeare & Next-Char Pred. & LSTM      & 80 & 422{,}615 \\
EMNIST      & Image Class.    & CNN       & 62 & 814{,}255 \\
\bottomrule
\end{tabular}
\caption{Statistics for datasets used in the evaluation.}
\label{tab:datasets}
\end{table}

\noindent
\textbf{Baseline policies.} We compare \sysname against three baselines.
(1)~\emph{FedAvg (single-tier)}~\cite{mcmahan2017fedavg}: the standard single-tier aggregation
in which all sites synchronize with one global aggregator every round. This represents the
communication structure of existing federated learning based geo-distributed training.
(2)~\emph{PowerTrip}~\cite{powertrip}: the state-of-the-art time-to-accuracy
optimizer, which dynamically selects which sites participate and how many participate but retains single-tier aggregation.
(3)~\emph{Static Hierarchy (Optimal)}: a hierarchy whose clustering, synchronization frequency,
and depth are fixed to the best configuration found by exhaustive offline search under the
\textit{initial} power and network conditions.

\noindent
\textbf{Power and communication profiling.} We model power availability by assigning each site a static power fraction drawn uniformly from $[0.1, 1.0]$. This value represents the available fraction of a site's maximum potential computational throughput. Following prior work~\cite{powertrip}, we assume an inverse linear relationship between execution time and available power for compute-bound machine learning workloads; a site's effective training time is its ideal time divided by its power fraction. For network profiles, each site is assigned a geographical distance drawn uniformly from 100 to 10{,}000 miles, with bandwidth scaling inversely proportional to distance. The model update size is configured to 1.5\,GB. While real-world inter-datacenter bandwidths and compute capacities may be larger in absolute magnitude, these emulated parameters preserve the fundamental ratio between computation and communication costs. Consistent with state-dependent accelerator power models~\cite{Kandiah}, we use normalized 
communication-to-compute and idle-to-compute power ratios 
of $\alpha = 0.30$ and $\beta = 0.15$, respectively.

\noindent
\textbf{Evaluation metrics.} Our primary metric is \emph{energy-to-accuracy}. This is the total energy consumed by the training sites to reach the target accuracy. It includes the compute, communication, and idle energy of all participating sites (\autoref{sec:single_tier}). We also report \emph{time-to-accuracy}. This is the wall-clock time required to reach the target accuracy. We use this metric to verify that energy savings do not degrade system performance.

\subsection{End-to-End Energy Efficiency}
\label{sec:eval_inefficiency}

\begin{figure}[t]
    \centering
    \includegraphics[width=\columnwidth]{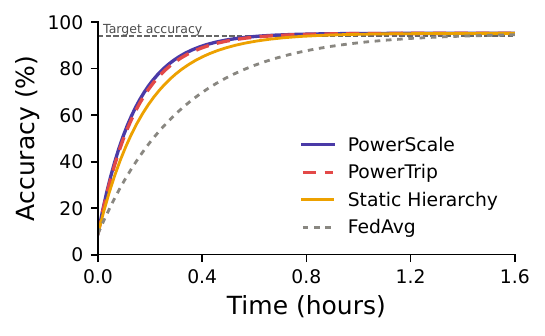}
    \caption{Test accuracy over training time for \sysname and baseline policies.}
    \label{fig:accuracy_vs_time}
\end{figure}

\begin{figure}[t]
    \centering
    \includegraphics[width=\columnwidth]{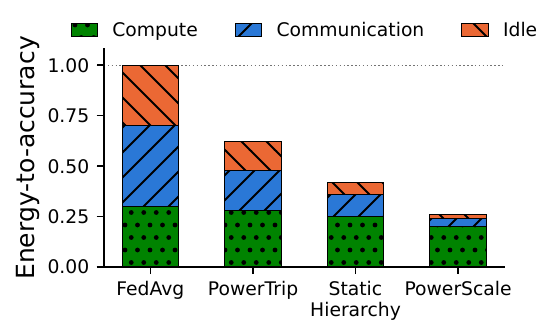}
    \caption{Energy-to-accuracy and energy component breakdown for \sysname and baseline policies. Each normalized to FedAvg.}
    \label{fig:energy_breakdown}
\end{figure}

We first evaluate whether \sysname's hierarchy improves end-to-end training efficiency. The
two figures in this section show complementary views of the same experiment.
\autoref{fig:accuracy_vs_time} plots test accuracy on the $y$-axis against wall-clock time
on the $x$-axis; the dashed horizontal line marks the target accuracy. This view captures
whether an energy-saving policy delays convergence. \autoref{fig:energy_breakdown} then
plots the energy required to reach that same target, normalized to FedAvg, with each
bar decomposed into compute, communication, and idle energy.

The convergence curves show that \sysname reaches the target accuracy in nearly the same
time as PowerTrip, while both substantially outperform FedAvg. Thus, the hierarchy does
not obtain its energy savings by slowing training down. The energy breakdown shows the
corresponding efficiency gain: \sysname consumes $3.9\times$ less total energy than
FedAvg and about $2.4\times$ less than PowerTrip. Because PowerTrip is the strongest
single-tier baseline in this comparison, the gap between PowerTrip and \sysname emphasizes
the value of adding intermediate aggregation. We discuss \sysname's time- and energy-to-accuracy against the static hierarchy policy in \autoref{sec:eval_clustering}.

\noindent
\textbf{Where the savings come from.} The stacked bars in \autoref{fig:energy_breakdown}
separate useful training work from coordination overhead. Compute energy changes modestly
across policies, since each method must perform enough local training to reach the same
target. The large reductions are instead in the overhead terms of Eq.~\ref{eq:energy_model}.
\sysname reduces communication energy by roughly $10\times$ relative to FedAvg by replacing
per-site WAN transmissions with per-cluster pre-aggregated updates. It also reduces idle
energy by roughly $15\times$ by limiting synchronous waiting to regional clusters and using
asynchronous aggregation across clusters.

\medskip
\noindent
\textbf{Key point.} \emph{\sysname preserves time-to-accuracy while reducing energy
by localizing communication and idle waiting.}

\subsection{Synchronization Modality}
\label{sec:eval_sync}

\autoref{sec:sync_modality} presented four synchronization configurations. We evaluate all four holding clustering and
frequency at \sysname's defaults, to isolate the effect of the synchronization mode.

\begin{figure}[t]
    \centering
    \includegraphics[width=\columnwidth]{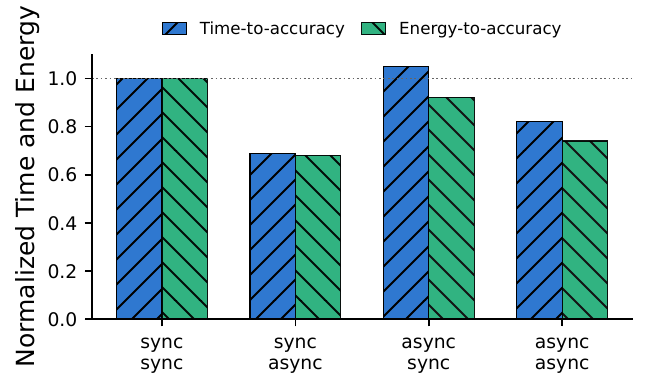}
    \caption{Energy-to-accuracy and time-to-accuracy (normalized to sync-sync) by synchronization mode.}
    \label{fig:sync_modality}
\end{figure}

\autoref{fig:sync_modality} places the four synchronization configurations on the $x$-axis and reports time-to-accuracy
(blue) and energy-to-accuracy (green) on the $y$-axis, each normalized to Sync--Sync. The
dotted line at 1.0 represents the fully synchronous baseline, and lower bars are better.

\noindent
\textbf{Sync-Async.}
Changing only the outer tier from synchronous to asynchronous reduces time-to-accuracy by 31\%
and energy-to-accuracy by 32\%. This result directly supports \sysname's Sync--Async
design. A synchronous outer tier
forces every cluster to wait for the slowest cluster before global progress can continue;
asynchronous aggregation removes that WAN-scale barrier. Retaining synchronous aggregation
inside each cluster preserves a coherent cluster model where the participants have already
been grouped to have similar completion times (\autoref{sec:dynamic_clustering}). 

\noindent
\textbf{Async-Sync.} Removing the cluster-tier barrier allows faster sites
to advance without waiting for slower members. But it also introduces intra-cluster
drift that requires 5\% more time to reach the target accuracy. The
per-round energy saving is small because intra-cluster variance is already
low, so the additional rounds yield only 8\% energy savings over Sync-Sync.

\noindent
\textbf{Async-Async.} Removing barriers at both tiers reduces time and
energy relative to Sync-Sync, but drift accumulates at both levels.
Compared to Sync-Async, convergence requires 14\% more time and energy
increases by 6\%.

\medskip
\noindent
\textbf{Key point.} \emph{Removing the global tier waiting barrier cuts both time and energy by roughly one third while preserving synchronous intra-cluster training.}

\subsection{Clustering Policy}
\label{sec:eval_clustering}

Clustering determines both communication and idle energy: it controls which
links carry synchronization traffic and how much completion-time variance
each synchronous group contains. We evaluate three decisions from
\autoref{sec:dynamic_clustering}: which clustering algorithm to use, which
signals to cluster on, and whether the grouping should adapt during training.

\begin{figure}[t]
    \centering
    \begin{subfigure}[t]{0.49\columnwidth}
        \centering
        \includegraphics[width=\linewidth]{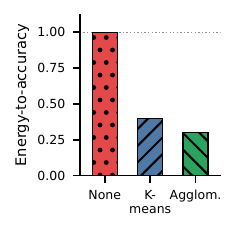}
        \caption{Algorithm}
        \label{fig:clustering_algorithms}
    \end{subfigure}\hfill
    \begin{subfigure}[t]{0.49\columnwidth}
        \centering
        \includegraphics[width=\linewidth]{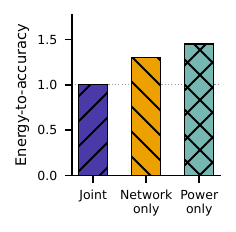}
        \caption{Criterion}
        \label{fig:clustering_criterion}
    \end{subfigure}
    \caption{Energy-to-accuracy for clustering algorithms (normalized to the no-clustering baseline) (a) and clustering criteria (normalized to Joint) (b).}
    \label{fig:clustering_policy}
\end{figure}

\noindent
\textbf{Clustering algorithms.}
As argued in \autoref{sec:dynamic_clustering}, K-means is most appropriate when the cluster count and
geometric structure are largely known in advance, whereas agglomerative clustering is the
better fit for a hierarchy that must be re-clustered online as power and network conditions change.
To evaluate these policies, \autoref{fig:clustering_algorithms} compares the two against a no-clustering baseline (PowerTrip's single-tier
aggregation) placed on the $x$-axis. The
$y$-axis reports energy required to reach the target accuracy, normalized to the no-clustering baseline.

K-means substantially reduces energy (by 60.3\%)
relative to single-tier aggregation by confining most traffic to shorter intra-cluster paths and
limiting straggler waiting to a smaller group of sites. Agglomerative clustering further reduces the
energy by 70.2\% compared to no clustering because its merge tree
supports incremental repair of cluster assignments without re-solving from
scratch. Thus, agglomerative reduces energy consumption by 33.2\% compared to K-means. 
\sysname therefore uses agglomerative clustering as its default.

The purpose of the experiment is not to claim that K-means
is a competitive runtime policy for \sysname; rather, it is to show that once a hierarchy must
adapt online, agglomerative clustering is the better mechanism.

\noindent
\textbf{Clustering criterion.} Next, we isolate the signals used to form clusters. The design in
\autoref{sec:dynamic_clustering} combines network proximity, which controls communication cost,
with available power, which controls site completion time. We compare this joint criterion
against network-only and power-only baselines.

\autoref{fig:clustering_criterion} places the three criteria on the $x$-axis and reports energy
to reach the target accuracy on the $y$-axis, normalized to joint clustering.
Network-only clustering consumes 29.8\% more energy and power-only clustering consumes 51.2\% more energy 
compared with joint clustering. Network proximity alone keeps transfers local, but it cannot
prevent a low-power site from pacing the synchronous inner loop. Incorporating power into the
feature space forms clusters whose members are better matched in both transfer and compute
time, reducing the waiting accumulated by faster sites. Conversely, the power-only
result reflects the complementary risk identified in the design: balancing compute without
preserving locality can place distant sites in the same cluster and increase communication
cost. \sysname therefore uses the joint criterion as its default.

\begin{figure}[t]
    \centering
    \begin{subfigure}[t]{\columnwidth}
        \centering
        \includegraphics[width=0.9\columnwidth]{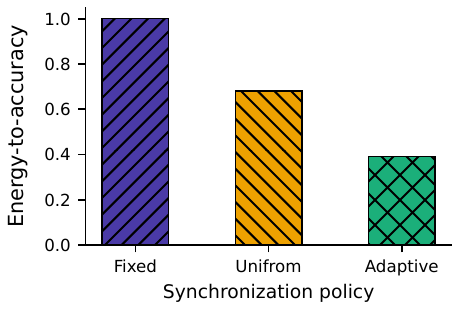}
        \caption{Energy-to-accuracy}
        \label{fig:frequency_ablation}
    \end{subfigure}
    \vspace{-0.4em}
    \begin{subfigure}[t]{\columnwidth}
        \centering
        \includegraphics[width=0.9\columnwidth]{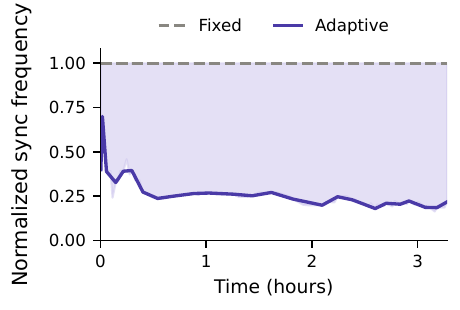}
        \caption{Synchronization frequency}
        \label{fig:sync_timeline}
    \end{subfigure}
    \caption{Energy-to-accuracy under three frequency policies (a) and global synchronization frequency over training progress (b). Both normalized to the fixed frequency policy. }
    \label{fig:frequency_policy}
\end{figure}

\noindent
\textbf{Dynamic vs. static clustering.} We compare full \sysname (dynamic re-clustering) against
a static baseline where the initial clustering is held fixed throughout training. 

\autoref{fig:accuracy_vs_time} and \autoref{fig:energy_breakdown} show
that dynamic re-clustering reduces energy-to-accuracy by 38.1\% and
time-to-accuracy by 22.5\% over a static hierarchy. A fixed grouping
cannot respond to power shifts, so sites whose available power drops
continue to stall their clusters and accumulate idle overhead.

\medskip
\noindent
\textbf{Key point.} \emph{Clustering is substantially better than no clustering, regardless of algorithm. Joint power-and-network clustering outperforms either criterion alone. Dynamic
re-clustering is essential under realistic, time-varying power.}

\subsection{Synchronization Frequency}
\label{sec:eval_frequency}

The final policy ablation evaluates how often clusters exchange updates
with the global tier. Following \autoref{sec:agg_frequency}, we compare
three frequency policies. The fixed policy applies the same synchronization
frequency to all clusters throughout training. The uniform policy applies
the same frequency to all clusters within each round but allows that
frequency to change over time. The adaptive policy assigns a per-cluster 
frequency proportional to each
cluster's global-to-local communication cost ratio and modulates it over
time based on observed convergence progress.

\autoref{fig:frequency_ablation} places these policies on the $x$-axis
and reports energy-to-accuracy on the $y$-axis, normalized to the fixed
policy. The adaptive policy reduces the energy required to achieve a given accuracy by 42.3\% compared with the uniform policy and by 60.7\% relative to the fixed baseline. The adaptive policy performs best because it accounts
for both spatial heterogeneity across clusters and temporal diminishing
returns over training.

\autoref{fig:sync_timeline} illustrates the temporal distribution of energy savings throughout the training process. The
$x$-axis is training time, while the $y$-axis reports effective synchronization
frequency. Higher
values therefore indicate more frequent coordination. The dashed gray line is the fixed
policy, while the
purple line is \sysname's adaptive policy. The adaptive policy synchronizes 2.43$\times$ more frequently in
early training than in late training, while the fixed baseline stays flat.
The shaded region between the two curves is the coordination the fixed
policy continues to perform but \sysname avoids: it widens as training
matures, because later updates are incremental and further global
synchronization adds little accuracy.

\medskip
\noindent
\textbf{Key point.} \emph{\sysname's adaptive frequency policy reduces energy by 60.7\% over a fixed
baseline by synchronizing less as the model converges and assigning each
cluster a frequency proportional to its communication cost.}

\subsection{Generalizability and Hierarchy Depth}
\label{sec:eval_general}

\begin{figure}[t]
    \centering
    \begin{subfigure}[t]{0.49\columnwidth}
        \centering
        \includegraphics[width=\linewidth]{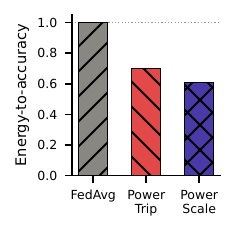}
        \caption{Shakespeare}
        \label{fig:generalizability_shakespeare}
    \end{subfigure}\hfill
    \begin{subfigure}[t]{0.49\columnwidth}
        \centering
        \includegraphics[width=\linewidth]{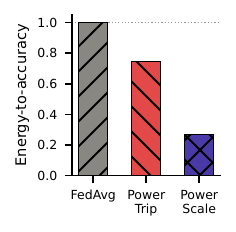}
        \caption{EMNIST}
        \label{fig:generalizability_emnist}
    \end{subfigure}
    \caption{Energy-to-accuracy on Shakespeare (a) and EMNIST (b), normalized to FedAvg.}
    \label{fig:generalizability}
\end{figure}

\noindent
\textbf{Cross-dataset generalizability.} We repeat the end-to-end
comparison on Shakespeare (LSTM) and EMNIST (CNN) to determine whether
\sysname's savings extend beyond CIFAR-10 and ResNet-18.
\autoref{fig:generalizability_shakespeare} and
\autoref{fig:generalizability_emnist} report energy-to-accuracy for the
three policies, normalized to FedAvg. The targets are 46\% next-character
accuracy for Shakespeare and 75\% classification accuracy for EMNIST.

On Shakespeare, \sysname uses 1.65$\times$ less energy than FedAvg and
1.15$\times$ less than PowerTrip. On EMNIST, the reductions are
3.70$\times$ over FedAvg and 2.75$\times$ over PowerTrip. The savings are
larger on EMNIST because its larger model updates make communication a
greater share of the single-tier baseline's total energy, which is
exactly the overhead the hierarchy removes. The policy ordering is
unchanged across both workloads, confirming that the benefit extends to
sequence prediction and image classification.

\begin{figure}[t]
    \centering
    \includegraphics[width=0.8\columnwidth]{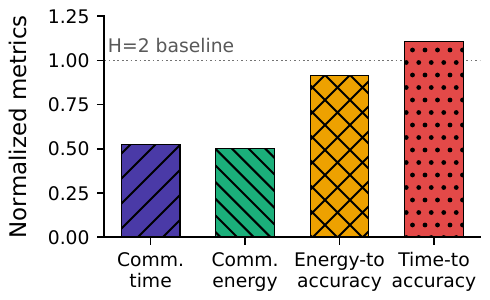}
    \caption{Effect of hierarchy depth at 100 sites. Each metric is normalized to $H{=}2$.}
    \label{fig:depth_crossover}
\end{figure}

\noindent
\textbf{Hierarchy depth.} The design adopts $H{=}2$ as the default and
treats depth optimization as out of scope (\autoref{sec:composed_policy}).
 Nevertheless, we compare $H{=}2$ and $H{=}3$ on the 100-site scale to
quantify the trade-off introduced by another aggregation tier.
\autoref{fig:depth_crossover} reports communication time, communication
energy, energy-to-accuracy, and time-to-accuracy, each normalized to $H{=}2$.

The third tier reduces communication time by 47.4\% and communication
energy by 49.8\%, lowering total energy-to-accuracy by 8.6\%. These savings come with
slower convergence: $H{=}3$ reaches the target in 35 rounds rather than
32, increasing time-to-target by 10.4\%. This is the depth trade-off
identified in \autoref{sec:composed_policy}: another tier consolidates
long-haul traffic but adds an aggregation boundary and accumulates
additional model drift. At 100 sites the 8.6\% energy reduction does not
justify the 10.4\% time penalty and the added system complexity, so
$H{=}2$ remains the default. The communication savings grow with the
number of cluster aggregators transmitting over the WAN, so deeper
hierarchies become more attractive at larger scales. Characterizing that
crossover requires depth runs across a range of site counts and remains future work.

\section{Related Work}
\label{sec:related}
\noindent
\textbf{Communication-Efficient Distributed Training.}
The challenge of communication overhead in distributed training is well-established~\cite{dean2012large, konecny2016federated}. Prior work focuses on reducing communication volume through gradient quantization~\cite{alistarh2017qsgd}, sparsification~\cite{aji2017sparse}, and parameter-efficient fine-tuning methods like LoRA~\cite{hu2022lora} and QLoRA~\cite{dettmers2023qlora}. Systems like DiLoCo~\cite{douillard2023diloco} and Photon~\cite{sani2024photon} leverage these techniques for distributed LLM training. These approaches optimize \emph{what} is sent; \sysname optimizes \emph{how} it is sent by restructuring the communication topology.

\noindent
\textbf{Hierarchical Federated Learning.}
HFL introduces intermediate aggregation layers between edge devices and a central server~\cite{luo2025deep, liu2020clientedgecloud}. Prior work addresses mobile and edge constraints such as heterogeneous devices~\cite{li2020fedprox}, privacy~\cite{geyer2017differentially}, and client--edge--cloud communication~\cite{zhang2024communication}. \sysname adopts the structural intuition of intermediate aggregation but addresses a different trade-off: it clusters data center sites by both network cost and available power, then minimizes energy-to-accuracy under WAN-bound communication rather than optimizing an edge deployment's battery use, privacy, or server load.

\noindent
\textbf{Adaptive Synchronization.}
Techniques such as local SGD~\cite{stich2018local} and adaptive aggregation frequency~\cite{chen2023synchronize, agarwal2021accordion} reduce communication by performing multiple local updates between synchronizations. Buffered asynchronous aggregation similarly mitigates stragglers by aggregating updates asynchronously in batches~\cite{nguyen2022fedbuff}. \sysname extends these ideas to a multi-tiered setting: clusters synchronize frequently within regions, while individual clusters adapt their costly inter-cluster synchronization rate to network cost and convergence progress. Its Sync-Async option applies asynchronous aggregation only at the outer tier, retaining synchronous regional aggregation where it limits local divergence.

\noindent
\textbf{Resource-Constrained AI Training.}
Prior work addresses power and resource constraints in AI training~\cite{patterson2021carbon, wu2022sustainable}. SkyPilot~\cite{yang2023skypilot} provisions GPU resources across clouds, optimizing for cost. Cross-region training~\cite{strati2024mltraining} analyzes distributed training across cloud regions. PowerTrip~\cite{powertrip} optimizes the power-communication trade-off in flat topologies. \sysname extends this line of work by introducing hierarchy to simultaneously optimize for time-to-accuracy and energy efficiency.

\noindent
\textbf{Energy-Efficient ML.}
EcoLearn~\cite{ecolearn} minimizes the carbon footprint of federated learning through carbon-aware client selection, and Green AI~\cite{schwartz2020green} advocates for energy-efficient training methods broadly. Zeus tunes job- and GPU-level configurations to navigate the energy--performance trade-off for individual DNN training jobs~\cite{you2023zeus}. NetStorm~\cite{wan2024netstorm} uses hierarchical communication to reduce WAN contention in distributed training. In contrast, \sysname makes the communication hierarchy itself an energy optimization decision: it dynamically clusters power-heterogeneous sites and adapts inter-cluster synchronization to minimize total energy-to-accuracy in geo-distributed data center training.

\vspace{-0.3cm}
\section{Conclusion}
\label{sec:conclusion}
This paper identifies and addresses the energy inefficiency of single-tier
aggregation in geo-distributed, power-constrained ML training. We introduce
\sysname, a system that utilizes hierarchical aggregation to reduce energy through
three policies: a Sync-Async modality that confines synchronous
waiting to low-latency regional clusters, agglomerative clustering on a
joint network-proximity and power-availability criterion, and an
adaptive synchronization frequency policy that tapers global synchronization as
training matures. \sysname reduces energy by $3.9\times$ relative to the single-tier aggregation baseline while matching or slightly improving
time-to-accuracy. Future work will investigate non-IID data distributions,
optimal hierarchy depth selection, and composition with gradient compression techniques.

\balance

\end{document}